Technical report

# Tracking and tracing in the UK: a dynamic causal modelling study

Karl J. Friston[1], Thomas Parr[1], Peter Zeidman[1], Adeel Razi[2], Guillaume Flandin[1], Jean Daunizeau[3], Oliver J. Hulme[4,5], Alexander J. Billig[6], Vladimir Litvak[1], Cathy J. Price[1], Rosalyn J. Moran[7] and Christian Lambert[1]

[1]*The Wellcome Centre for Human Neuroimaging, University College London, UK*
[2]*Turner Institute for Brain and Mental Health & Monash Biomedical Imaging, Monash University, Clayton, Australia*
[3]*Institut du Cerveau et de la Moelle épinière, INSERM UMRS 1127, Paris, France*
[4]*Danish Research Centre for Magnetic Resonance, Centre for Functional and Diagnostic Imaging and Research, Copenhagen University Hospital Hvidovre, Kettegaard Allé 30, Hvidovre, Denmark.*
[5]*London Mathematical Laboratory, 8 Margravine Gardens, Hammersmith, UK*
[6]*Ear Institute, University College London, UK*
[7]*Centre for Neuroimaging Science, Department of Neuroimaging, IoPPN, King's College London, UK*

**E-mails**: *Karl Friston, k.friston@ucl.ac.uk; Thomas Parr, thomas.parr.12@ucl.ac.uk; Peter Zeidman, peter.zeidman@ucl.ac.uk; Adeel Razi, adeel.razi@monash.edu; Guillaume Flandin, g.flandin@ucl.ac.uk; Jean Daunizeau, jean.daunizeau@googlemail.com; Oliver Hulme, oliverh@drcmr.dk; Alexander Billig, a.billig@ucl.ac.uk; Vladimir Litvak, v.litvak@ucl.ac.uk; Rosalyn Moran, rosalyn.moran@kcl.ac.uk; Cathy Price, c.j.price@ucl.ac.uk; Christian Lambert, christian.lambert@ucl.ac.uk*

## Abstract

By equipping a previously reported dynamic causal model of COVID-19 with an isolation state, we modelled the effects of self-isolation consequent on tracking and tracing. Specifically, we included a quarantine or isolation state occupied by people who believe they might be infected but are asymptomatic—and only leave if they test negative. We recovered *maximum posteriori* estimates of the model parameters using time series of new cases, daily deaths, and tests for the UK. These parameters were used to simulate the trajectory of the outbreak in the UK over an 18-month period. Several clear-cut conclusions emerged from these simulations. For example, under plausible (graded) relaxations of social distancing, a rebound of infections within weeks is unlikely. The emergence of a later second wave depends almost exclusively on the rate at which we lose immunity, inherited from the first wave. There exists no testing strategy that can attenuate mortality rates, other than by deferring or delaying a second wave. A sufficiently powerful tracking and tracing policy—implemented at the time of writing (10th May 2020)—will defer any second wave beyond a time horizon of 18 months. Crucially, this deferment is within current testing capabilities (requiring an efficacy of tracing and tracking of about 20% of asymptomatic infected cases, with less than 50,000 tests per day). These conclusions are based upon a dynamic causal model for which we provide some construct and face validation—using a comparative analysis of the United Kingdom and Germany, supplemented with recent serological studies.

Technical report

Contents

# Introduction

This is the third in a series of technical reports that use dynamic causal modelling to explain and predict the current outbreak of COVID-19. The first report described an enhanced compartmental model based upon a factorisation of latent or hidden states generating timeseries data, such as new cases and daily deaths (Friston et al., 2020a). This model was concerned with an outbreak in a single region, parameterised with an effective population size. The second report assembled several models of a single region, coupled by population flux between regions, to model the pandemic in the United States of America (Friston et al., 2020b). The focus of this multi-region model was on the genesis of second waves and a key, mechanistic, distinction between rebounds due to premature relaxation of social distancing and second waves due to loss of immunity. The basic conclusions were that a devolved social distancing strategy—that was sensitive to local metrics—predicted better outcomes than a national or federal strategy. In this report, we return to the model of a single region or country and look more closely at strategies in terms of surveillance; specifically, the role of testing, tracking and tracing.

The efficacy of contact tracing programs is now the focus of several modelling initiatives (Aleta A et al., 2020; Ferretti et al., 2020; Giordano et al., 2020; Gurdasani and Ziauddeen, 2020; Hellewell et al., 2020; Keeling et al., 2020), whose conclusions depend upon the form of the models used. Models that include social distancing and isolation of infected contacts suggest that a 'find', 'track', 'trace' and 'isolate' (FTTI) policy can ameliorate morbidity (Giordano et al., 2020; Kretzschmar et al., 2020). The terms FTTI and "tracking and tracing" will be used interchangeably throughout.

To address the efficacy of FTTI, we equipped the dynamic causal model (DCM) with a further location state; namely, a state of self-isolation or quarantine. People entered the state when experiencing symptoms or awaiting a PCR test[1]. They remained isolated for seven days unless they received news that the test was negative. This construction accommodates the mechanistic process by which FTTI operates. In other words, the agenda behind tracking and tracing is to isolate people who are infected before they become contagious. This allows one to move back in time and pre-empt the reproduction of the virus in the population. However, to do this, it is necessary to identify people who are asymptomatic, thereby enriching or enhancing the probability that targeted testing will identify infected individuals. We operationalise this tracking and tracing strategy in terms of its *efficacy*. Here, efficacy is defined as the probability that I will be offered a test by an FTTI programme if I am infected and asymptomatic. Under this parameterisation, an ineffective tracking and tracing renders this probability zero[2]. Conversely, an efficiency of 100% means that if I am infected and asymptomatic, I will certainly be tested. Clearly, for a large population, high levels of efficacy may not be attainable; however, lower levels may be sufficient to either suppress the reproduction rate of viral transmission (Aleta A et al., 2020) or defer the emergence of any second wave until an efficacious programme of vaccination is in place (or effective treatments have been established).

To model different aspects of testing and surveillance, we had to carefully parameterise testing along a number of dimensions. To do this, we assume that there was a small, time-dependent probability of being tested on any given day. This *testing* probability was modelled in terms of a constant *baseline*, a testing component *sensitive* to the prevalence of infection in the population and a *sustained* component following the first wave. This sustained component was modelled as proportional to the level of herd immunity acquired after successive waves of infection. Having parameterised the testing probability, the *selectivity* of testing was parameterised in terms of the probability of being tested if infected, relative to not being infected. Finally, if I am infected but asymptomatic, then the probability of being tested is supplemented with a *track and trace* component—that could start at the beginning of the outbreak, or any subsequent time. This may sound a rather involved parameterisation; however, it is a minimal model needed to generate the number of positive cases reported, given the latent prevalence of infection. This follows because the number of positive cases depends not only on the probability of being tested but whether I am more likely to be tested if I am infected (e.g., I work in a care home) or not (e.g., I have been selected at random by a screening survey).

Please see Figure 1 and Table 1 for a brief review of the model (and the appendices for the parameterisation of self-isolation and testing). With this model and its parameters in place, one can now fit the model to empirical data until the present day. Crucially, because the parameters of this model do not change with time, they can be used to forecast the future trajectory, under various adjustments to the testing parameters. The following conclusions foreshadow the results of these simulations:

- There is no plausible parameterisation of the model that would or could permit a flareup or rebound of the first wave following a relaxation of social distancing measures. This is under the qualified

---

[1] This assumes 100% compliance with self-isolation protocols, which is itself a key issue. Please see Webster, R.K., Brooks, S.K., Smith, L.E., Woodland, L., Wessely, S., Rubin, G.J., 2020. How to improve adherence with quarantine: rapid review of the evidence. Public Health 182, 163-169.

[2] More precisely, the probability that I would have been tested irrespective of my special status.

- assumption that social distancing continues to be operating in the way it is modelled—and inferred—on the basis of the empirical evidence to date. In short, provided there is a graded, social distancing response to the prevalence of infection in the population, there will be no rebound in the weeks following the peak of the first wave.

- A second wave (distinct from a rebound of the first wave) is inevitable. The timing of the second wave depends almost exclusively on the rate at which immunity is lost. In other words, under the assumption that infection confers immunity—and that the immunity lasts for a given period—the period of immunity determines the timing of the second wave. This second wave is mechanistically distinct from a fluctuation of, or rebound from, the first wave.

- In the absence of an effective therapeutic intervention (i.e. vaccine or treatment), there is no social distancing or surveillance strategy that will have any material impact on the total number of deaths accumulated from the onset of an outbreak to an idealised endemic equilibrium. However, certain strategies can defer waves of infection, affording more time to develop treatments that lower morbidity. Specifically, tracking and tracing can defer expression of the second wave beyond a time horizon, after which vaccination or other therapeutic interventions will render it innocuous. In short, the mechanism by which strategic interventions operate is not eliminating the infection[3] but slowing it down sufficiently, so that its pathogenicity is dissolved by viral mutation, vaccination, or therapeutic advances. Here, we assume a time horizon of 18 months.

- The most efficacious strategy for deferring a second wave is tracking and tracing. Furthermore, the logistic requirements are within current capabilities. The same is not true of the first wave. In other words, it would not have been possible to completely suppress the first wave with tracking and tracing because one would have had to have identified (nearly) every infected, asymptomatic person in the country and this would have required over a million tests a week.

- Although, in principle, it is mathematically possible to defer the first wave, one would require either a very small population or a very large testing capacity. Furthermore, the efficacy of tracking and tracing would have to be high, i.e., around 80%.

- The differences between the United Kingdom and Germany are eminently explainable under a dynamic causal model. As might have been anticipated, Germany has a greater propensity to test; however, this testing is substantially less selective for infected individuals than in the UK. Furthermore, the tracking and tracing component due was *less* evident than in the United Kingdom. This means that pressure is put on other parameters to explain the remarkably low fatality rates in Germany. It appears that—or it looks as if—the Germany population had demographic or host factors that rendered it more resistant to infection. In short, the explanation for the reduced fatalities in Germany probably lies in their population density, clinical surveillance, and management, not on tracking and tracing.

---

[3] But see Alexander F. Siegenfeld and Yaneer Bar-Yam, Eliminating COVID-19: A community-based analysis, arXiv:2003.10086 (March 23, 2020).

In what follows, we will look at the results of simulations that license the above conclusions and unpack these conclusions quantitatively, with a special emphasis on the mechanisms and processes leading to different outcomes.

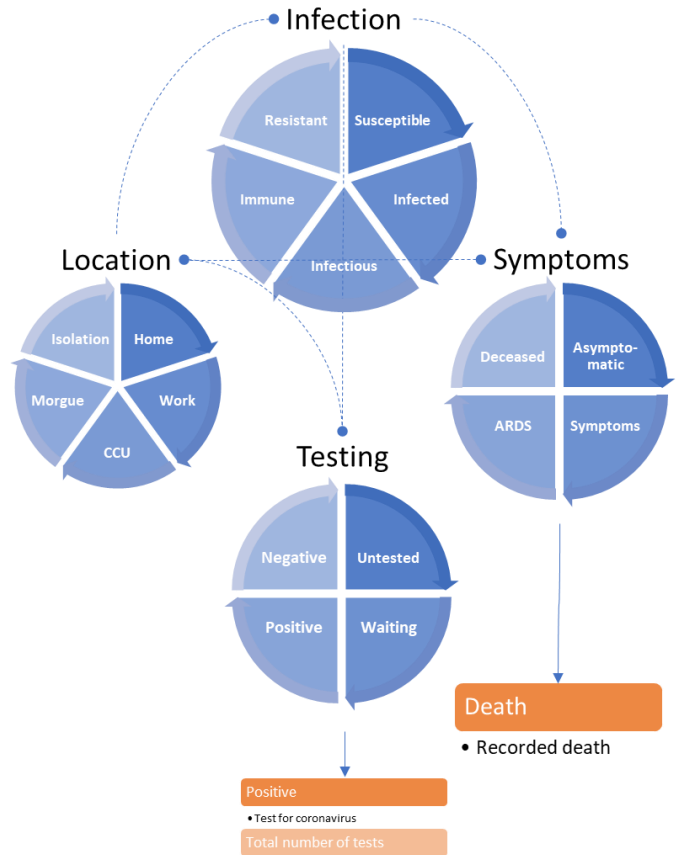

Figure 1

**The LIST model**: This schematic summarises the LIST (*location*, *infection symptom* and *testing*) generative model used for the following simulations. This model is formally similar to that described in (Friston et al., 2020b). Here, it has been augmented with an additional location state (*isolation*) to model people who are self-isolating because they think they may be infectious (within their home or elsewhere). Note that in this model there are no absorbing states. In other words, one can leave any state within any of the four factors. For example, one only occupies the state of being deceased (or testing positive and negative) for a day and then moves to asymptomatic (or untested) on the following day. This ensures that the total population is conserved, i.e., probability mass is conserved in terms of the ensemble density. Furthermore, it enables the occupancy of various states to be interpreted in terms of the rate of daily expression. The blue boxes correspond to states or compartments. The states within any factor are mutually exclusive, where the factors embody the factorial form of this compartmental model. In other words, every individual in the population has to be in one of several possible states that are characterised in terms of four factors or attributes. The orange boxes represent the observable outputs that are generated by this dynamic causal model, in this instance, daily reports of positive tests, daily tests and deaths.

TABLE 1

Parameters of the epidemic (LIST) model and priors, $N(\eta, C)$

(NB: prior means are for scale parameters $\theta = \exp(\vartheta)$)

| Number | Parameter | Mean | Variance | Description |
|---|---|---|---|---|
| 1 | $\theta_n$ | 4 | 1 | Number of initial cases |
| 2 | $\theta_r$ | 1/3 | 1/256 | Proportion of resistant cases |
| 3 | $\theta_N$ | 66 | 0 | Population size (millions) |
| **Location** | | | | |
| 4 | $\theta_{out}$ | 1/3 | 1/256 | Probability of going out |
| 5 | $\theta_{sde}$ | 1/32 | 1/256 | Social distancing threshold |
| 6 | $\theta_{cap}$ | 16/100000 | 1/16 | CCU capacity threshold (per capita) |
| **Infection** | | | | |
| 7 | $\theta_{Rin}$ | 4 | 1/256 | Effective number of contacts: home |
| 8 | $\theta_{Rou}$ | 48 | 1/256 | Effective number of contacts: work |
| 9 | $\theta_{trn}$ | 1/3 | 1/256 | Transmission strength |
| 10 | $\theta_{inf} = \exp(-\frac{1}{\tau_{inf}})$ | $\tau_{inf} = 4$ | 1/16 | Infected period (days) |
| 11 | $\theta_{con} = \exp(-\frac{1}{\tau_{con}})$ | $\tau_{con} = 4$ | 1/16 | Infectious period (days) |
| 12 | $\theta_{imm} = \exp(-\frac{1}{\tau_{imm}})$ | $\tau_{imm} = 16$ | 0 | Period of immunity (months) |
| **Symptoms** | | | | |
| 13 | $1 - \theta_{dev} = \exp(-\frac{1}{\tau_{inc}})$ | $\tau_{inc} = 5$ | 1/256 | Incubation period (days) |
| 14 | $\theta_{sev}$ | 1/128 | 1/256 | Probability of ARDS |
| 15 | $\theta_{sym} = \exp(-\frac{1}{\tau_{sym}})$ | $\tau_{sym} = 8$ | 1/256 | Symptomatic period (days) |
| 16 | $\theta_{rds} = \exp(-\frac{1}{\tau_{rds}})$ | $\tau_{rds} = 10$ | 1/256 | ARDS period (days) |
| 17 | $\theta_{fat}$ | 1/3 | 1/256 | ARDS fatality rate: CCU |
| 18 | $\theta_{sur}$ | 1/8 | 1/256 | ARDS fatality rate: home |
| **Testing** | | | | |
| 19 | $\theta_{ttt}$ | 1/10000 | 1 | Efficacy of tracking and tracing |
| 20 | $\theta_{sen}$ | 1/10000 | 1 | Sensitivity of testing |

Technical report| 21 | $\theta_{exp}$ | 1/10000 | 1 | Sustained testing |
| 22 | $\theta_{bas}$ | 8/10000 | 1 | Baseline testing |
| 23 | $\theta_{tes}$ | 1 | 1/16 | Selectivity of testing infected people |
| 24 | $\theta_{del} = \exp(-\frac{1}{\tau_{del}})$ | $\tau_{del} = 2$ | 1/256 | Delay in reporting test results (days) |

**Secondary sources** (Huang et al., 2020; Kissler et al., 2020; Mizumoto and Chowell, 2020; Russell et al., 2020; Verity et al., 2020; Wang et al., 2020) and:

- https://www.statista.com/chart/21105/number-of-critical-care-beds-per-100000-inhabitants/
- https://www.gov.uk/guidance/coronavirus-COVID-19-information-for-the-public
- http://www.imperial.ac.uk/mrc-global-infectious-disease-analysis/COVID-19/

These prior expectations should be read as the effective rates and time constants as they manifest in a real-world setting. For example, a four-day period of contagion is shorter than the period that someone might be infectious (Wölfel et al., 2020)[4], on the (prior) assumption that they will self-isolate, when they realise they could be contagious. The priors in this table differ slightly from previous reports, with some (e.g., clinical) parameters becoming more precise and others (e.g., testing) being relaxed. This reflects the fact that more data are now available; thereby requiring a rebalancing of the priors, with respect to the likelihood of the data. Although the scale parameters are implemented as probabilities or rates, they are estimated as log parameters, denoted by $\vartheta = \ln \theta$.

# Dynamic causal modelling

Figure 1 provides a schematic that summarises the dynamic causal model used for subsequent inference and simulations. This model can be regarded as a factorial extension of conventional (compartmental) epidemiological models (Friston et al., 2020a). The factorial aspect means that there are several attributes in play when trying to model the causes of mortality and morbidity[5]. Specifically, it considers the location, infection status, symptomatology, and testing status of any individual in a population. This means that for each factor, there is a certain probability of finding someone in a particular state. Movement from one state to another is parameterised in terms of transition probabilities or rate constants. For example, the probability that I will stay in a state of being *infectious* can be parameterised in terms of the expected number of days that I am contagious. Crucially, the transitions among states within each factor depend upon the states of other factors. These dependencies are denoted by the dashed lines. For example, the probability that I will move from a state of having no symptoms (*asymptomatic*) to symptoms depends upon whether I am *infected* or not. Note that separating the latent or hidden causes of observable data in this way means that it is possible to be infected but have no symptoms – and *vice versa*. This model is formally the same as previous models (Friston et al., 2020b); however, we have introduced a fifth location state called *isolation*. This state is entered whenever I have *symptoms* or am waiting to find out whether I test *positive*. The key mechanism—that compels me to enter *isolation*—is a tracking and tracing (FTTI) program that alerted me to the possibility of being infected prior to developing symptoms. In this quarantined isolation, I will remain for a given period (seven days), unless I receive notice that I have tested negative following a PCR test. Please see the appendices for a formal parameterisation of these contingencies and how testing data are generated.

---

[4] Shedding of COVID-19 viral RNA from sputum can outlast the end of symptoms. Seroconversion occurs after 6-12 days but is not necessarily followed by a rapid decline of viral load.

[5] The factorial aspect of dynamic causal models like these enables one to exploit conditional dependencies during model inversion. This has two fundamental advantages. First, the models are generally more comprehensive than conventional epidemiological models because they can accommodate more states (e.g., thousands as opposed to tens). Second, conditional dependencies licence the use of variational procedures (e.g., mean field approximations) that dominate Bayesian modelling and machine learning outside epidemiology. Variational procedures are orders of magnitude more efficient than conventional sampling (e.g., Metropolis Hastings) procedures. This means one can handle many more parameters and optimise the model using Bayesian model comparison (a.k.a. structural learning).

With this model in place, one can use standard variational procedures to fit any data at hand (Friston et al., 2007). Here, we used the daily reports of new (positive) cases and deaths from Johns Hopkins University[6] and supplemented this with data from the UK on the total number of tests performed[7]. The inversion of this model takes about a minute on a personal computer, enabling one to generate posterior estimates of the parameters and accompanying trajectories of hidden states. Figure 2 shows the results of this kind of analysis for timeseries data at the point of writing (10th May 2020). The left panels show the data (dots) superimposed upon a posterior predictive density. This density is a probabilistic statement about the most likely outcomes under the model. Here, it is summarised in terms of the posterior expectation or most likely outcome and 90% Bayesian credible intervals (blue lines and shaded areas, respectively). The parameters upon which these predictions are based are shown in the lower right panel. The outcomes can either be expressed in terms of daily rates or cumulative outcomes over time; for example, the cumulative deaths over a six-month period (as shown on the upper right panel). Notice that these posterior predictive densities cover the past and the future. In other words, they are generated from parameter estimates that do not change in time. This means that one can regard these results as a best fit to the observed data to date. Alternatively, they can be regarded as a forecast of the future. These results suggest that we are nearly halfway down the decline in daily death rates, following the peak (in early April). Crucially, because we have a generative model underneath these data, we can also generate data that have not yet been observed.

---

[6] Available from https://github.com/CSSEGISandData/COVID-19. These timeseries were smoothed with a Gaussian kernel to suppress spurious fluctuations at the weekends.

[7] Available from https://github.com/tomwhite/covid-19-uk-data. These timeseries were smoothed with a Gaussian kernel to suppress spurious fluctuations at the weekends.

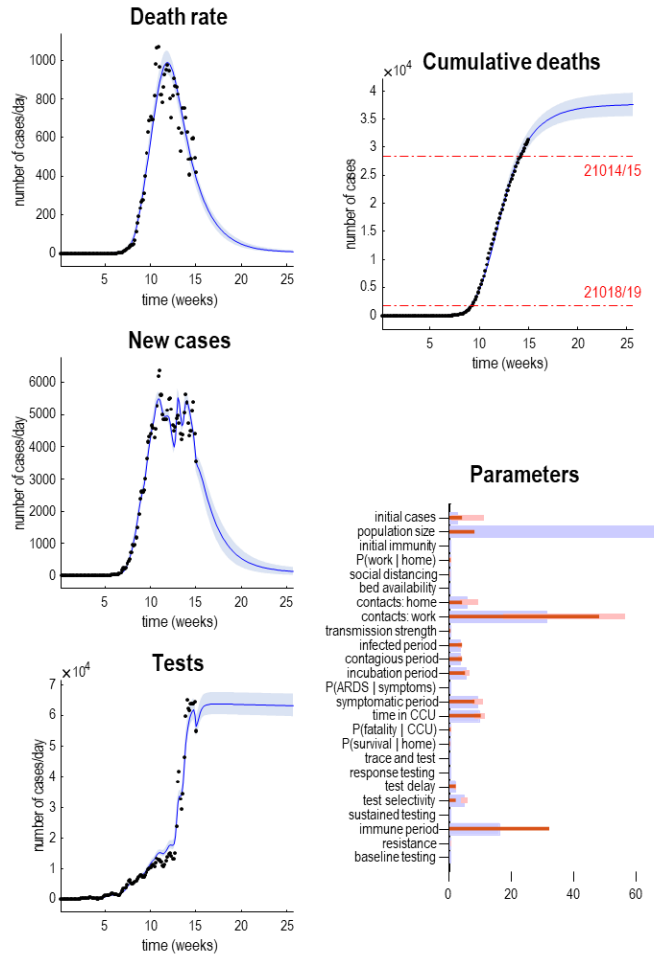

**FIGURE 2**

**Posterior predictive densities**: The panels on the left show the posterior predictive densities over some key outcomes. Here, daily death rates, new positive cases and tests performed. The blue lines represent the posterior expectation and the shaded areas the 90% Bayesian credible intervals. The black dots correspond to empirical data used to fit the model and estimate posteriors over the model parameters (i.e., the transition probabilities or rate constants in Table 1). The lower right panel reports the posterior parameter densities in terms of their posterior expectation (blue bars) and 90% credible intervals (pink bars). The red bars correspond to the prior expectations. Please see Table 1 for complete specification of the prior densities. The upper right panel shows the cumulative deaths expected under these parameters. The two dashed lines are for reference and correspond to yearly mortality rates for seasonal influenza (from 2014/15 and 2018/19). Time zero corresponds to 25th January 2020.

Figure 3 provides two examples of this, in terms of the effective reproduction rate (R) and the prevalence of immunity in the left and right panels, respectively. The prevalence of immunity (a.k.a. herd immunity) is interesting because it is potentially measurable, if we had serological testing of sufficient sensitivity and specificity (Winter and Hegde, 2020; Yong et al., 2020). Were such data to become available, they could be used to improve the posterior estimates of the parameters and shrink uncertainty about the trajectory of seroprevalence (Vespignani et al., 2020) or immunity (Kissler et al., 2020).

Notice that the reproduction rate is treated here as an outcome. This is an important conceptual point. The reproduction rate is not a cause of fatality—it is a measure or consequence of the underlying causes. This can be computed from the changes in the prevalence of infection and the expected duration of being contagious (see appendices). Here, the reproduction rate starts at just under three and then falls quickly at the onset of social distancing to about 0.6. In the future, the model predicts it will gently rise as herd immunity is lost and may ultimately foreshadow a second wave (see below). The yellow line corresponds to the best available estimates of the reproduction ratio based on hierarchical Bayesian Regression, using specific known covariates (regressors) corresponding to different stages of lockdown (Flaxman et al., 2020).

These estimates[8] are the kind of numbers used to currently guide governmental policy in the UK. They can be regarded as the best estimates from state-of-the-art curve fitting, sometimes supplemented with careful consideration of delays and other (e.g., contact) data (Jarvis et al., 2020). The key observation here is that these estimates necessarily depend upon data that have already been observed. In other words, they summarise the recent past. This can be seen by comparing the yellow line with the blue line in Figure 3. The real-time estimates afforded by dynamic causal modelling (blue line) evince a more nuanced decline that precedes the sharp drop in conventional estimators (yellow line). This speaks to the potential advantage of using estimates of latent states to furnish real-time or instantaneous estimates of the reproduction rate (as opposed to retrospective curve fitting or Bayesian regression estimators).

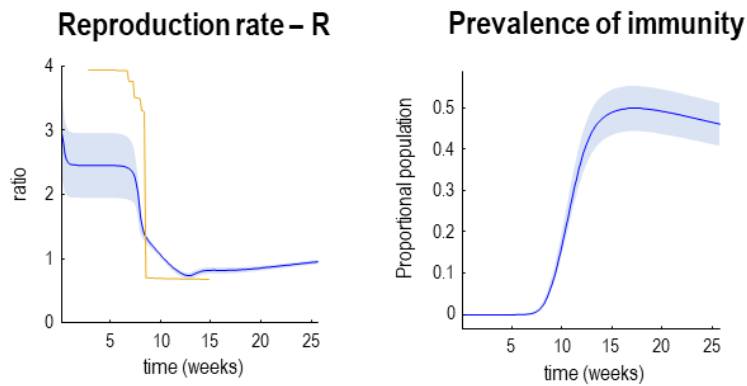

**FIGURE 3**

**Reproduction rates and herd immunity**: This figure supplements the previous figure with posterior estimates of the reproduction rate (see appendices) and the prevalence of immunity (a.k.a. herd immunity). Again, the lines represent posterior expectations and the shaded areas 90% credible intervals. The yellow line in the right panel depicts estimates based upon a Bayesian regression model. These are the kind of estimates used to inform government policy. Please see main text for further discussion.

Figure 4 reproduces the results above (in the upper panels) and supplements these outcomes with the latent causes or hidden states that correspond to the factors in Figure 1. Here, we have shown the course of the

---

[8] Available from https://mrc-ide.github.io/covid19estimates/#/download.

Technical reportTechnical report

pandemic over 18 months, as opposed to a six-month period. This illustrates the basic anatomy of the pandemic with an initial first wave, followed by a second wave some 36 weeks (i.e., nine months) later. The timing of this second wave depends strongly upon the rate at which immunity is lost. In these simulations, a 16-month period of immunity was assumed[9].

Focusing on the initial outbreak (i.e., first wave), we can see the effects of social distancing as manifest in a very small probability of being found at work during the period of lockdown (blue line in the *location* panel). This coincides with a large number of people self-isolating (about 60% at its highest) during this period (purple line in the *location* panel). Under this model, we are currently experiencing the relaxation of social distancing, with a partial return to the pre-pandemic probability of being found at work. However, the world to which we return differs from that before the lockdown. This is because a substantial number of people have acquired immunity, in virtue of being infected (whether or not they show any subsequent symptoms). The acquisition of herd immunity (a.k.a. population immunity) is depicted by the yellow line in the *infection* panel. Although there are no data that inform these estimates, equivalent data from Germany are starting to appear. We will return to this later. Notice that a substantial proportion of the UK population (about 38%) have been estimated by the model to be resistant[10]. In other words, they have geographical or host factors that render them unlikely to participate in the pandemic. For example, they may be isolated geographically[11] or may have genetic or developmental factors that preclude infection and viral shedding. In terms of *symptoms*, the most prevalent expression of the pandemic is in terms of symptoms that accompany an increase in self-isolation—and pre-empt a smaller subset of people who go on to develop a severe syndrome (e.g., acute respiratory distress) from which they may not recover. The *testing* panel shows the progressive increase in people waiting to be tested (blue line) that subsumes people who are subsequently positive and negative. Initially, the number of negative tests is about twice the number of positive tests, but this ratio increases as the number of infected people in the population declines.

These predicted trajectories are inferred from the data under the prior assumptions inherent in the model. Crucially, this model incorporates social distancing as a process that responds to other latent causes or hidden states. Here, social distancing (i.e., the probability of leaving the *location* state *home*) is modelled as a soft threshold function of the prevalence of *infection* in the population (Friston et al., 2020b). This allows the social distancing threshold to the estimated. In other words, social distancing is built into the model as a cause of the epidemiology and ensuing data. Figure 5 illustrates the predictions of social distancing, in terms of the probability of staying at home and its underlying cause; namely, the prevalence of infection. There is a remarkable correspondence between predicted social distancing behaviour and the times at which the government announced social distancing and lockdown measures—and their relaxation. The question now is: what would happen to these trajectories under different monitoring or testing policies over the next few months?

---

[9] This is an intermediate period of immunity that falls between a rapid loss of immunity, characteristic of betacoronaviruses responsible for the common cold—and an enduring immunity over many years that is conferred by most vaccinations.

[10] Note that this posterior estimate is constrained by fairly precise priors, with a standard deviation of 1/16.

[11] See https://www.esade.edu/itemsweb/wi/research/ecpol/EsadeEcPol_Insigth6_Exit_Strategy.pdf for a discussion of isolation in terms of social networks and green zones.

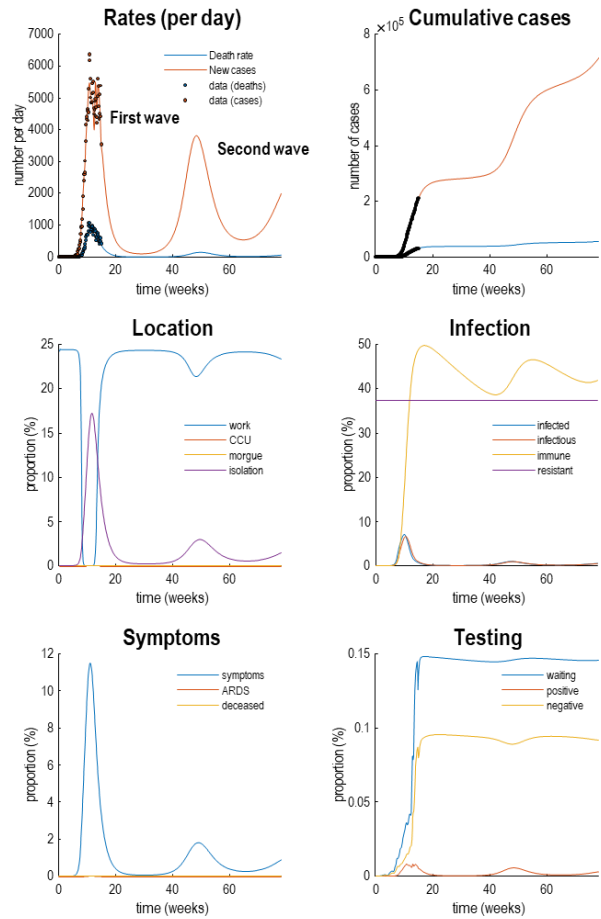

**FIGURE 4**

**Latent causes**: This figure shows the same data (with dots) and posterior expectations (solid lines) as in the previous figure over an 18-month period. Here however, it is supplemented with the underlying latent causes or expected states in the lower four panels (the first state in each factor has been omitted for clarity: i.e., *home*, *susceptible*, *asymptomatic,* and *untested*). These hidden states generate the outcomes in the upper two panels. The solid lines are colour-coded and correspond to the states of the four factors in Figure 1. For example, under the *location* factor, the probability of being found at work declines steeply from about 20% to 0% at the onset of the outbreak. At this time, the probability of isolating oneself rises to about 15% during the peak of the pandemic. After about five weeks, the implicit social distancing starts to relax and slowly tails off, with accompanying morbidity (in terms of symptoms) and mortality (in terms of death rate). As herd immunity (yellow line in the *infection* panel) declines the prevalence of infection accelerates to generate a second wave that peaks at about 48 weeks.

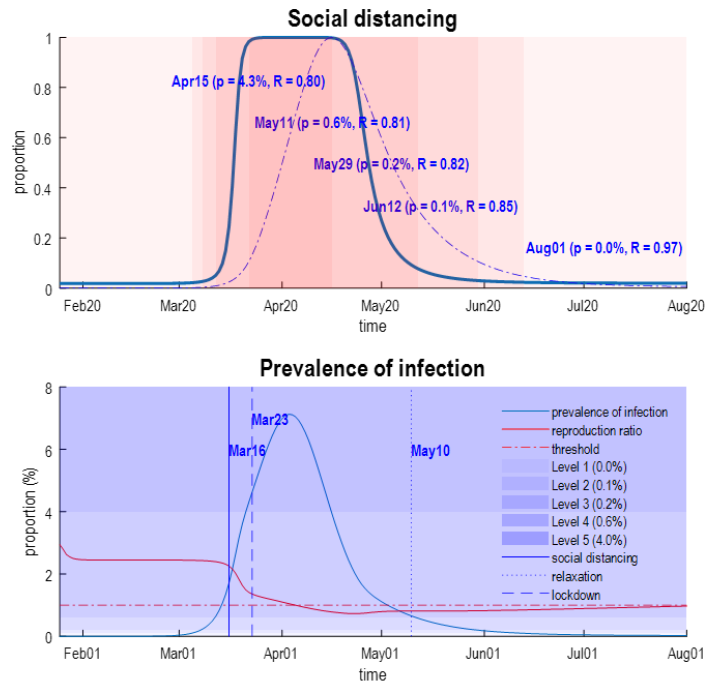

FIGURE 5

**Social distancing**: This figure illustrates the predictions of social distancing based upon the hidden states in the previous figure. The upper panel quantifies social distancing in terms of the probability of remaining at home, as a function of time (solid line). This is supplemented with the daily death rate, normalised to its maximum value (broken line). The shaded areas discretise this continuous characterisation of social distancing into five levels. These levels correspond to threshold crossings of the prevalence of infection. This prevalence is shown as the blue line in the lower panel. For illustration, the thresholds defining the five levels of social distancing are .1%, .2%, .6% and 4% (for reference, the posterior expectation of social distancing threshold in Table 1 was 3.5%). Threshold intervals are highlighted with shaded areas. The prevalence of infection (blue line) is supplemented with the accompanying reproduction ratio (red line). This is a function of the underlying prevalence; namely, the proportional rate of change in the number of infected people. The broken horizontal line corresponds to a reproduction rate of one. The vertical lines show when social distancing was introduced (on March 18), when the UK went into lockdown (on March 23) and the beginning of relaxation (on May 10). Note that the model did not know about these dates: social distancing here is the best explanation for how new cases and daily deaths are generated—not an explanation for when various interventions were introduced. Transitions from one level to the next are annotated in the upper panel in terms of the date, the prevalence of infection at that time and the estimated reproduction rate. For example, on April 15, this model predicts that the prevalence of infection was 4.3% with an R of 0.8. According to this analysis, social distancing will end, as we currently know it, on June 12[12]. At this time, the prevalence of infection will have fallen to 0.1%; however, the reproduction rate will have increased slightly—and will continue to rise with the loss of population immunity. Under an assumed period of immunity of 16 months, it will approach one by August. Noted, under these (posterior) predictions, there will still be a nontrivial death rate due to COVID-19 when social distancing is relaxed completely. Note that previous analyses suggest that a return to a pre-pandemic level of social discourse is unlikely. In other words, the probability of being found at *work* will return to slightly lower levels. This probability stands in for all forms of social distancing measures (e.g., wearing facemasks).

---

[12] This is the same date predicted in our first report, focusing on the outbreak in London: "*By June 12, death rates should have fallen to low levels … and social distancing will no longer be a feature of daily life*" *Friston, K., Parr, T., Zeidman, P., Razi, A., Flandin, G., Daunizeau, J., Hulme, O., Billig, A., Litvak, V., Moran, R., Price, C., Lambert, C., 2020a. Dynamic causal modelling of COVID-19 [version 1; peer review: awaiting peer review]. Wellcome Open Research 5.*

Figure 6 provides an analytic answer to this question in terms of the effects on daily death rates—as a function of time—as testing parameters are adjusted[13]. Here, we evaluated the effect of enhancing testing and monitoring policies by adjusting the parameters that underwrite tracing and tracking (blue line), testing sensitivity (red line), testing delay (yellow line), testing selectivity for infected people (purple line) and, finally, the baseline probability of being tested (green line). The upper panel shows the effect on daily deaths when each of these parameters is increased by a scaling factor of one natural unit (i.e., 2.72). The key thing to observe is that the effect of changing these parameters itself changes over time. Here, we considered a period of 18 months; under the assumption that by the end of this period, there will be an effective vaccination program or other therapeutic advances.

There are two key things to note from this sensitivity analysis. First, the effect of any testing parameter on the first peak (before the vertical blue line) is much smaller than the effect on the second peak. This second peak emerges because of a loss of immunity, modelled here with an immune period of 16 months (see Table 1). The second thing to note is that the effects are biphasic in nature. For example, increasing baseline testing initially decreases death rates with both the first and second waves, but increases death rates after the waves peak. At first glance, this may seem counterintuitive; however, there is a simple explanation. This rests on the fact that any surveillance measure has the effect of delaying the spread of the virus (via self-isolation measures), such that the onset of successive waves of infection is suppressed and the peak is deferred or pushed into the future. In other words, increased surveillance affords the opportunity to reduce the spread of the virus, such that successive waves of infection are delayed and dispersed—along the lines of the 'curve flattening' notion. Indeed, this was the primary motivation for social isolation to avoid excess mortality due to a saturating clinical care capacity. However, in the absence of any limitations on critical care, surveillance, in and of itself, cannot attenuate the eventual spread of the virus throughout the population—it can only delay the spread. Metaphorically, this process can be imagined as rain falling from clouds. Eventually, the downfall will reach the sea. The only thing that one can do is to moderate the flow of water and mitigate against flood damage.

---

[13] In this and subsequent figures, we will drop credible intervals to avoid visual clutter. All of the quantities reported are equipped with credible intervals; however, here, our agenda is to illustrate the form of epidemiological trajectories and how they depend on various testing strategies. A proper risk analysis would include the conditional uncertainty that attends these trajectories.

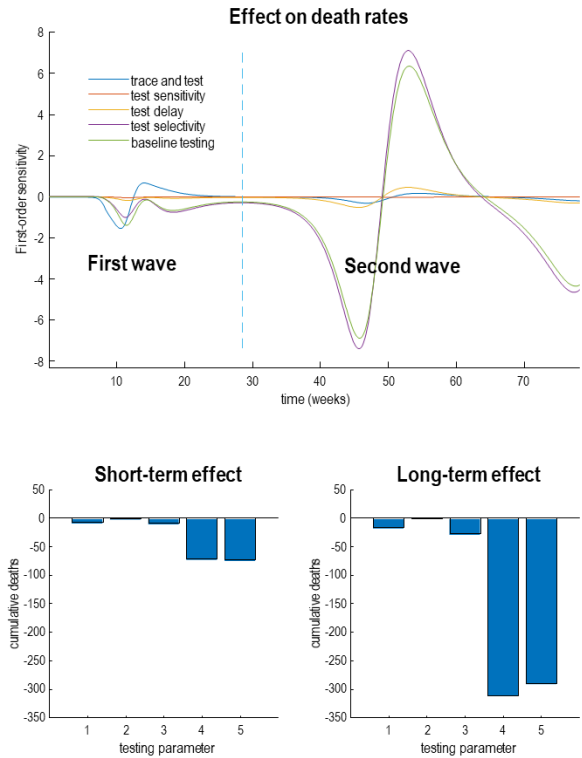

FIGURE 6

**Sensitivity analysis**: This figure illustrates the influence on death rates of various parameters that model diagnostic surveillance or testing. The upper panel shows the change in death rates with respect to the logarithm of the parameters controlling the efficacy of tracing and tracking (blue line), PCR testing in response to increasing levels of infection (red line), delay in reporting test results (yellow line), the selectivity for people who are infected (purple line) and baseline testing (green line). The lower panels show the cumulative changes over time. The lower left panel sums the changes over a 20-week period following the onset of the outbreak, while the lower right panel accumulates the changes over a period of 18 months. The key thing to observe is that the effect of changing testing or surveillance parameters is more marked during the second wave, relative to the first. Furthermore, the change in accumulated deaths, with respect to a unit change in log parameters, is very small (in the hundreds as opposed to the thousands). This reflects the fact that the effect of these testing parameters is to shift the curve, not to attenuate its amplitude.

Quantitatively, this key point is reflected in the overall number of lives that will be saved by enhancing one aspect of surveillance or another. The lower panels in Figure 6 show the cumulative number of lives saved under the five different parametric enhancements. As might be expected, increasing surveillance in various ways generally decreases the cumulative deaths; however, these effects are quantitatively very small: in the tens for an effect after the first wave (left panel) and in the hundreds after the second wave (right panel). This suggests that the utility of enhanced surveillance (e.g., tracing and tracking) can only be manifest if the second wave is pushed sufficiently far into the future, such that it is rendered innocuous through vaccination or other therapeutic interventions.

This is illustrated in Figure 7 which simulates the trajectories that one might expect when increasing the efficacy of tracking and tracing (see the appendices for how efficacy is parameterised). This figure uses the same format as Figure 4 but reproduces trajectories under increasing levels of tracking and tracing. In brief, one can see that there is hardly any effect on the first wave in terms of social distancing (*location*), prevalence of infection (*infection*), or morbidity (*symptoms*). However, the peak of the second wave is shifted progressively later in time, until it disappears beyond the 18-month time horizon. These simulations, as noted above, used a loss of immunity with a time constant of 16 months. This may be a somewhat pessimistic estimate of the rate at which we lose immunity; however, it clearly demonstrates the utility of tracking and tracing under this scenario. In summary, as the efficacy of tracking and tracing increases, the second wave is progressively deferred, and the number of positive cases detected in the population rises. These effects are highlighted with the blue and orange arrows. So, what levels of tracking and tracing would be necessary to preclude a second wave within a time horizon of 18 months?

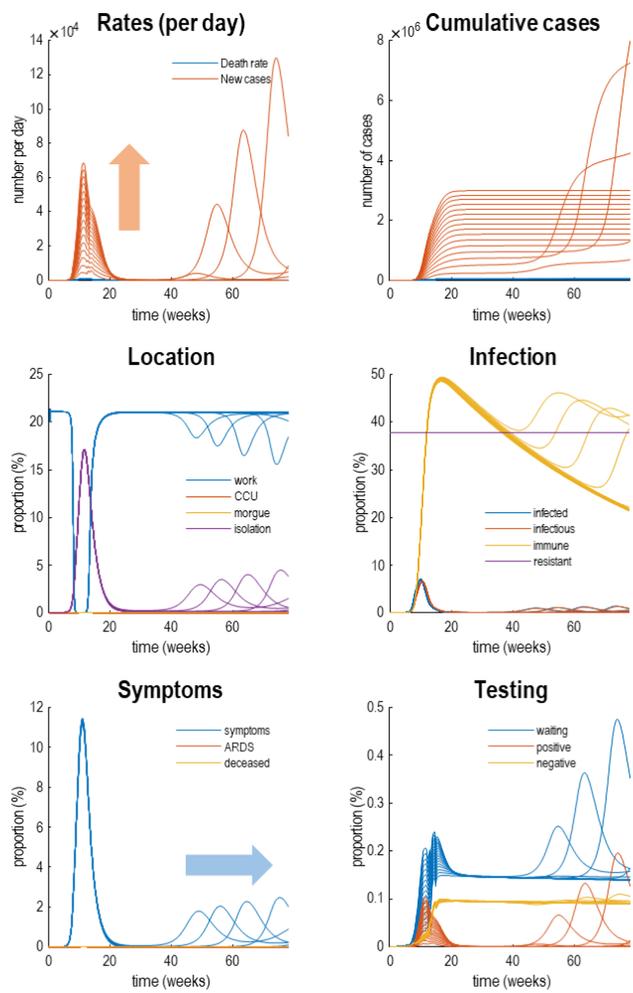

**FIGURE 7**

**Tracking and tracing**: This figure reports the results of simulations under different protocols for tracking and tracing. Specifically, we increased the probability of testing people who were infected but asymptomatic after the first wave.

This increase in efficacy was from 0 to 1 in 32 steps. The results are shown using the format of Figure 4, for every fourth step. The key insight from this figure is that as one increases the efficacy of tracking and tracing, the second wave is deferred or postponed beyond a time horizon (here, 18 months). At the same time, the total number of detected (positive) cases per day increases. These effects are summarised by the blue and orange arrows, respectively.

Figure 8 answers this question by plotting the cumulative deaths and peak testing rates as a function of the efficacy of tracking and tracing. These posterior predictions were based on increasing the efficacy of tracking and tracing from 0 to 1 in 32 steps—depicted every four steps in the previous figure. As the efficacy of tracking and tracing increases there is a marked reduction in cumulative deaths in the order of 10,000 people. This reflects the delay in the second wave (solid line). Quantitatively, it would be sufficient to have an efficacy of about 24% to defer the second wave until 18 months, under the prior assumptions of the model. According to these estimates, this would entail peak testing rates of less than 10,000 tests per day, well within the reach of current testing capacity.

The dotted lines show the corresponding predictions for a tracking and tracing strategy that was instantiated prior to the first wave. These results are interesting in the sense that they speak to what might have happened had the UK government pursued a tracking and tracing strategy at the onset of the pandemic. In principle, there was a potential to defer the first wave and thereby elude any deaths due to COVID-19. This is shown by the second arrow in the left panel of Figure 8. However, things are not quite that simple. In order to eliminate the first wave (ignoring additional measures such as restricting inbound travel), it would have been necessary to have an efficacy of tracking and tracing of about 80% or more. In other words, nearly everybody who was infected but asymptomatic would have to have been identified. A more realistic efficacy of 50% would have reduced deaths in the initial phases of the outbreak; however, this would have required peak testing rates beyond the capacity of a country like the United Kingdom.

This is illustrated by the dashed line in the right panel that surpasses an arbitrary threshold of 250,000 tests a day. In short, although a suppression strategy based on tracking and tracing is a theoretical possibility, it cannot be realised after the number of infected people exceeds testing capacity. It is interesting to speculate what this means for countries like South Korea and Singapore who have managed to elude a substantive first wave. In virtue of the fact that they have not acquired a meaningful herd immunity, they may have to maintain a high level of efficacy of FTTI, in conjunction with strict border controls and accompanying quarantine. From the perspective of the United Kingdom, which at the time of writing is likely past the peak of its first wave, the question is: do the same mechanics of surveillance apply to the second wave?

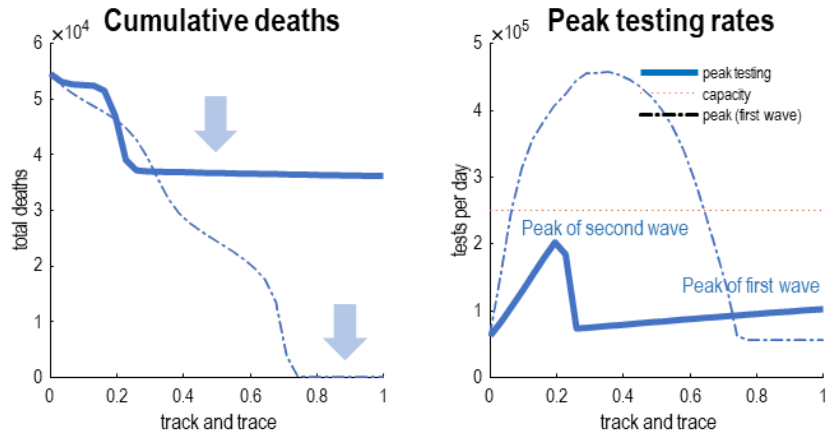

FIGURE 8

**Suppressing waves**: This figure summarises the results of the previous figure in terms of cumulative deaths after an 18-month period, as a function of the efficacy of tracking and tracing. The left panel shows the total number of deaths as a function of the efficacy of a FTTI protocol that starts after the first wave (solid line) or before the first wave (broken line). The equivalent results are shown in the right panel in terms of the requisite peak testing rates over the course of the outbreak. For an FTTI protocol that starts after the first wave, all efficacies require plausible peak testing rates. An efficacy of about 24% or greater sufficient to defer or delay the second wave until it can be rendered innocuous. The accompanying reduction in cumulative deaths is shown by the upper blue arrow. The lower blue arrow highlights the equivalent effect had tracking and tracing been implemented at the onset of the outbreak—and maintained at efficacy levels of over 70%. With this efficacy tracking and tracing, the first wave could have been delayed indefinitely, reducing deaths to negligible levels. However, with 50% efficacy the peak tests per day would have exceeded testing capacity. This capacity is illustrated by the red line in the right panel (here, 250,000 tests per day). In short, if one had a small country or exceedingly well developed FTTI resources, it would have been possible to eliminate the first wave; however, for a country like the United Kingdom, this may not have been an option. The discontinuity in peak testing rate (in the right panel) reflects the fact that under low levels of efficacy, the greatest testing rate is required at the second wave; however, if the second wave is deferred, peak testing coincides with the first wave.

The answer to this question is no. This is because the context in which the second wave manifests is very different from the first wave. This follows because of the acquisition of herd immunity, which means that the spread of the virus—in the run-up to the second wave—is substantially attenuated. In turn, this means that the requisite efficacy of FTTI is substantially smaller. This point is illustrated in Figure 9 (top left panel) by evaluating the predicted outcomes at the (18-month) time horizon under four scenarios. The first scenario was based upon the posterior estimates of the current testing parameters (*current*). The second scenario entailed an enhanced baseline testing (*enhanced testing*). The third scenario was an enhancement of selective testing (*selective testing*); namely, increasing the relative probability of testing infected people. Finally, we consider a FTTI strategy (*tracing*), in which the efficacy was increased to 25%. The upper panels show the posterior predictions as a function of time (left panel) and as a phase-space summary of the same trajectories (right). This way of detecting trajectories plots one outcome against another: i.e., plotting the daily rates of new cases against daily deaths.

The resulting trajectories illustrate the effects of various interventions. In brief, as we increase the rate or selectivity of testing, we shift the trajectories upwards. In other words, we increase the number of detected cases but with little effect on daily deaths. In contrast, the FTTI strategy suppresses the second wave and reduces daily deaths. The lower panel quantifies the endemic endpoint at the time horizon of 18 months. This is not an endemic equilibrium but stands in for the state of affairs at the point of a presumed vaccination or therapeutic intervention. One can see that the various testing strategies progressively reduce the daily death rates at this endpoint. For example, with an FTTI efficacy of 25%, the daily deaths due to COVID-19 are about 10 per day. This is roughly the number of people who are killed in road traffic accidents[14]. The number of tests at this time, are reasonably manageable (about 50,000 per day for the FTTI strategy). Crucially, the number of lives saved is reduced considerably under, and only under, FTTI. In this example, the elimination of the second wave would save about 16,000 lives. Notice that simply elevating the sensitivity or selectivity of testing has little effect on mortality rates. Only the FTTI strategy enables the early identification of infected individuals, their subsequent isolation and ensuing deferment of a putative second wave.

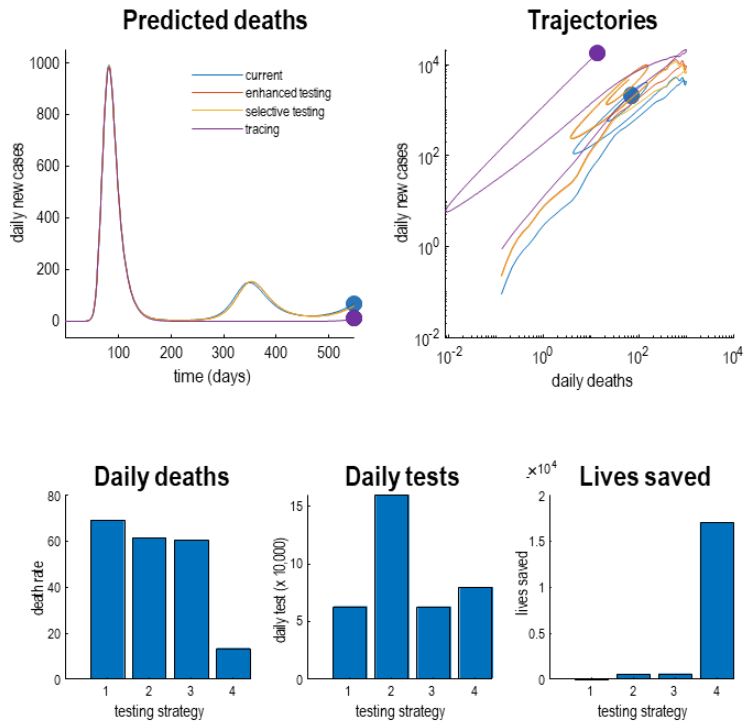

FIGURE 9

**Different testing strategies**: This figure reports simulations of what might happen under different testing or surveillance regimes. The upper panels show the simulated daily deaths predicted under four strategies. These include the current strategy based upon the posterior expectations of testing parameters. The red lines show the response to an enhanced baseline testing. The yellow lines simulate the outcomes under an increase in selective testing, while the purple lines illustrate the impact of tracking and tracing with an efficacy of 25%. For the enhanced testing, the posterior expectations of the log parameters were increased by a value of one (i.e., the scale parameters were multiplied by a natural unit: 2.72). The FTTI parameter was increased to 25%. The upper left panel plots the predicted deaths as a function of time for an 18-month period. The upper right panel shows the same data but plotted as a trajectory in a phase-space, spanned by daily deaths and reports of new cases. This illustrates the fact that, mathematically, the trajectories into the future correspond to flows towards an attracting orbit or set. In this instance, the systems have point attractors. However, here, we have assumed that a vaccine is available at 18 months, at which point the trajectories terminate in the filled circles (shown for the first and last testing scenarios). The lower three panels characterise this endpoint in terms of daily deaths, daily tests and total number of lives saved since the onset of the outbreak. It can be seen that, as might be anticipated, the successive testing enhancements reduce daily deaths. The most expensive strategy, in terms of daily tests, is the enhanced baseline testing, requiring 160,000 tests again. The remaining strategies require a more modest 70,000 tests per day. The most efficient and life-saving strategy is the implementation of tracking and tracing that could, under this model, save more than 15,000 lives. The filled circles in the upper panels denote the endpoints that, here, stand in for the endemic equilibria under the first and last strategies. Note that the tracking and tracing strategy is the only approach that materially decreases daily deaths, both over time and at the 18-month endpoint.

---

[14] https://www.gov.uk/government/collections/road-accidents-and-safety-statistics.

An important (if obvious) observation, implicit in this treatment, is that testing can be deployed in different ways with distinct agendas. Crucially, the only kind of testing that matters for saving lives is identifying those individuals who are infected before they can spread the virus. This is the *raison d'être* for FTTI, as opposed to simply increasing test rates. Increasing the baseline, sensitivity or selectivity of testing provides more precise data for epidemiological modelling and subsequent policy decisions; however, in and of itself, it will not have any material effect on the progression of the pandemic. Similarly, testing people who are symptomatic is too late from the point of view of isolating individuals who may become contagious—even if it allows people to return to work early. Clearly, all three agendas are important; however, it may be useful to consider (and model) testing in terms of its distinct aims; namely, to defer a second wave, to enhance epidemiological surveillance and to ease pressure on the economy and clinical care.

## A comparative analysis

The conclusions from the above modelling are clear. There is an imperative to instantiate (arguably re-instantiate) FTTI at modest levels of efficacy in the next few months—to preclude a second wave by delaying it. Furthermore, trying to maintain an early FTTI strategy at effective levels would have been logistically difficult. This begs the question: how has Germany managed to suppress its mortality rates, if it contended with the same kind of outbreak confronting the United Kingdom? To answer this question—at a fairly crude level—we repeated the dynamic causal modelling using daily new cases and deaths from Germany. The trajectories of these outcomes and their latent causes are shown in the upper and lower panels of Figure 10, respectively. In the absence of data on the total number of tests, we assumed a constant baseline testing (see Appendix).

The ensuing differences in the outcomes and latent states speak to what we already know. For example, despite having about the same number of people testing positive during the first peak, the mortality rates in Germany are about a quarter of those witnessed in the UK. The inferred surveillance and testing suggest that Germany started with a baseline testing rate, such that at any one time 0.1% of the population was waiting for their test results. The UK, conversely, accrued its testing capacity during the first wave and, according to these estimates, now exceeds the German testing rates. Despite increased testing in Germany, the number of people self-isolating was about half that in the UK (as estimated under this model), with less than 10% of the German population quarantining themselves at the peak of the pandemic. Furthermore, Germany's social distancing was less stringent and shorter, as reflected by the blue lines in the location panel of Figure 10. The infection panel is telling; in the sense that about 38% of the UK population are estimated to be resistant. However, this rises to about 58% of the German population. This is a marked difference suggesting either geographical or host factors may play an important role in the differential fatality rates. Indeed, when one examines the underlying posterior parameter estimates for the trajectories depicted in Figure 10, it becomes clearer how Germany and the UK differ.

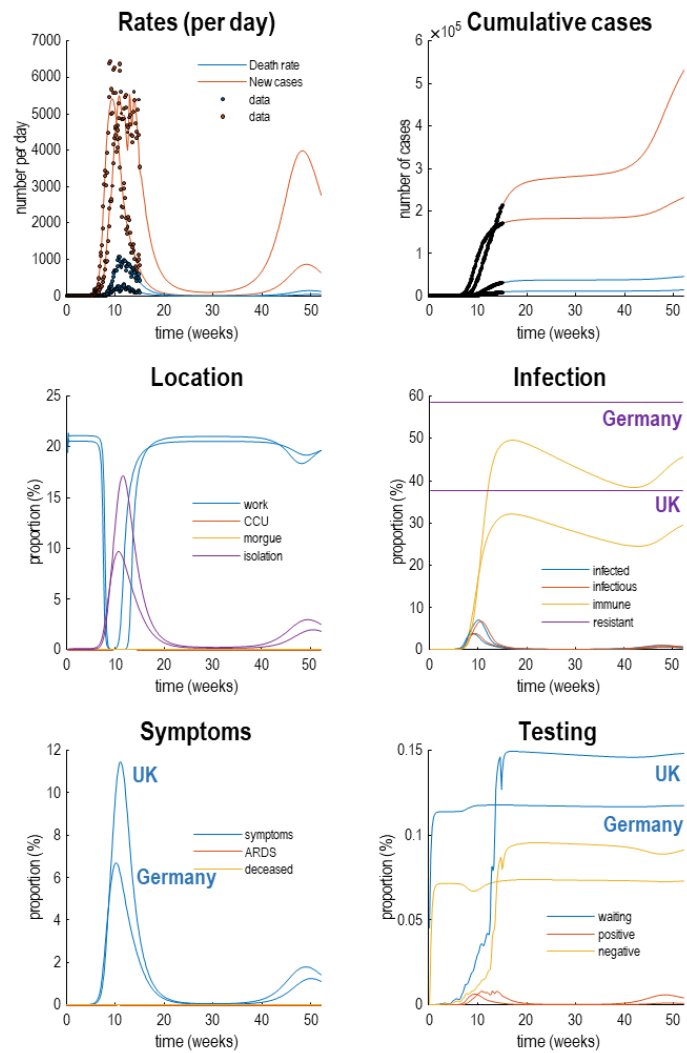

FIGURE 10

**Germany and the UK compared**: This figure shows the latent causes (lower panels) of observed and predicted outcomes (upper panels) for Germany and the UK, where time zero corresponds to 25 January 2020. The coloured pairs of lines in each panel use the same format as Figure 4 and refer to the two countries in question. The generative model provides a good account of the empirical data for both countries (black dots in the upper panels), with formally similar fluctuations in latent epidemiological states. However, there are some key quantitative differences. For example, the degree of self-isolation and social distancing is attenuated by roughly one half in Germany. This pertains both to the percentage of people self-isolating and the duration of social distancing at a societal level. This is also reflected in the lower prevalence of symptomatic individuals at the first (and second) peaks of infection. These differences are, in large part, due to the number of people who are susceptible to infection, as reflected in the proportion of people who are resistant (about 38% for the UK and 58% for Germany). The parameters that underwrite these trajectories are shown in the next figure.

Figure 11 shows the parameters with the greatest difference between Germany and the UK, in terms of the country specific estimates (upper panels) and the differences (lower panels). The parameters are shown in terms of log parameters (left panels) and the corresponding scale parameters (right panels). The scale parameters are nonnegative rate constants and probabilities, while the log parameters are simply the log transformed scale parameters. For clarity, only the 12 parameters with the greatest posterior difference are shown. They have been ranked such that the parameters on the left show the greatest difference (the parameters are labelled by the subscripts in Table 1). The key thing to take from this comparison is that there are marked differences between Germany and the UK, both in the testing parameters and the parameters pertaining to susceptibility and clinical surveillance. Indeed, the most marked difference is a fivefold increase in the sensitivity of German testing to the prevalence of infection. This testing is nearly 5 times less selective for infected people than in the UK. This lower selectivity means that testing is more targeted to those with symptoms in the UK compared to in Germany. This is consistent with what we know from the German approach relative to the U.K.'s approach.

The third largest difference is the number of people infected (*n*) at the beginning of the timeseries. The inference here is that Germany started (on 25th of January 2020) with about three times as many infected people as the United Kingdom. By virtue of the fact that Germany tested more sensitively but non-selectively from the onset of the outbreak, the sustained testing component (*exp*) is much smaller. Note also that the FTTI parameter (*ttt*) is also smaller. In other words, under this model there is no evidence that tracking and tracing in Germany was a key component of their surveillance program. The key parameters to note here are the substantial (about 50%) increase in the proportion of the German population that were resistant to infection and a (about 20%) decrease in the probability of fatality in critical care. Finally, the probability of developing severe symptoms when infected is slightly lower than in the UK. Note that these are all estimates based purely upon daily reports of new cases and deaths. They are not informed by any data relating to survival rates in critical care or clinical studies of susceptibility. Rather, they are the simplest explanation for the data at hand, where simplicity here means a minimum departure from our prior beliefs about how the data are generated (Penny, 2012).

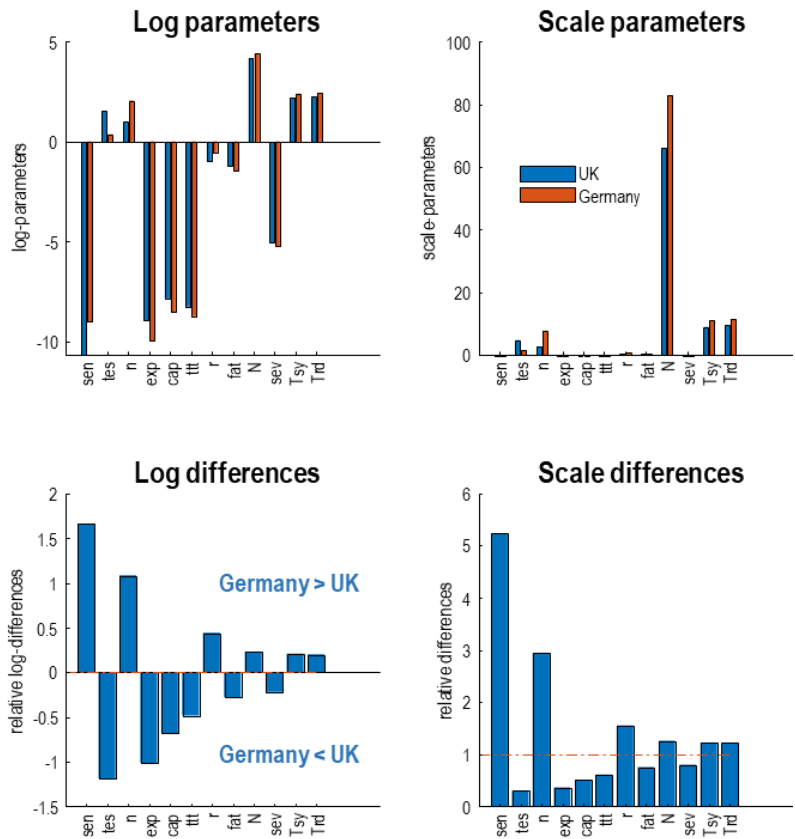

FIGURE 11

**A parametric comparison**: This figure shows the posterior expectations of the model parameters form Germany and the UK. The upper panels show the estimates for each country separately, while the lower panels show the differences. The left panels show the parameters in terms of their logarithmic form, while the right panels show the exponentiated (scale) parameters. In the upper panels, the UK parameters are in blue and the German parameters are in orange. The labels on the X axis correspond to the parameter subscripts in Table 1. The key thing to take from this figure is that the most marked (quantitative) differences between the two countries lies in the parameters pertaining to testing; namely, the sensitivity to the prevalence of infection, the selectivity of testing people who are infected and sustained testing after the first peak. Having said this, the efficacy of tracking and tracing is actually less in Germany—as estimated by the model—than in the UK, while Germany appears as if it has a more sensitive testing or surveillance program that is less selective for people who are infected. These testing parameter differences explain why the number of positive cases reported is about the same for United Kingdom and Germany, while the inferred number of people who are infected (and subsequently die) is much less in Germany (by a factor of roughly 4). The explanation, under this model, for this differential mortality lies in the clinical and management parameters. These include an increased number of resistant members of the population and a reduced fatality rate, when severely ill. Furthermore, there is a decrease in the probability of developing severe symptoms when infected.

The evidence here contradicts the hypothesis that Germany's relatively low fatalities are caused by a more vigorous testing programme. At one level, this hypothesis is naive because testing cannot cause morbidity. Test results and deaths are consequences of morbidity. As noted above, the only way that testing can affect fatalities is by delaying the spread of the virus—and this is most effective when infected but asymptomatic individuals are identified through tracking and tracing. A more plausible interpretation of these parameter estimates, and ensuing predictions is as follows:

The proportion of people testing positive who subsequently died in Germany is lower not because people who are infected are less likely to die but simply because Germany has tested more people[15]. These reduced mortality rates may reflect several factors: for example, a lower population density in Germany, demographics of the affected cohorts, differences in the use of post-mortem testing and variation in clinical practice. Under the current model, some of these may factors contribute to the increase in the inferred proportion of the population who are resistant (see conclusion). This is consistent with anecdotal evidence that the seeding of the outbreak in Germany was via skiers from Italy and, presumably, disseminated among a younger and healthier subset of the population[16]. The relative reduction in the probability of developing severe symptoms and subsequent fatality may reflect the clinical surveillance and management of symptomatic people. For example, anecdotal reports from respiratory physicians in Germany suggest a more prospective clinical management, with lower thresholds for admission to critical care. In contrast, much of the disease burden in the UK appears to have been managed in an elderly and vulnerable population in care homes. This cohort are unlikely to survive the rigours of intubation in an intensive care unit and their clinical management is necessarily more palliative. In short, Germany may have had to deal with a different kind of problem than that confronting the UK. In short, although German testing and clinical surveillance was more evident, the clinical surveillance mattered more in terms of mortality.

Clearly, this is purely speculation; however, in principle, it should be possible to evaluate the evidence for these speculative hypotheses when more detailed data become available. This speaks to one application of dynamic causal modelling to compare different models in terms of their evidence (Penny et al., 2004); for example, comparing a model of outcomes in Germany and the UK with, and without, country-specific differences in the parameters.

# Conclusion

The key conclusions from this kind of modelling are twofold. First, the notion of a flareup or rebound of infections is not supported by the evidence at hand. A popular conception of this rebound is akin to lowering the 'flood gates' too soon and being overwhelmed with a deluge of infections. However, this picture may be a false impression. If immunity lasts on the order of months, then there can be no flood because a sufficient proportion of the population have already been exposed to the virus. These people preclude a rapid spread of the virus through the population by acting as a retardant or buffer that suppresses the effective reproduction ratio. In other words, the first wave cannot flareup because it may have exhausted the substrate of susceptible individuals it needs to disseminate itself. This speaks to the second key

---

[15] Note that the implicit infection mortality implied by these estimates is based upon the inferred levels of infection not reported levels. In other words, reports of new cases are the data that are explained by the inferred prevalence that accommodates a bias towards (or away from) testing people who are infected.

[16] See https://www.telegraph.co.uk/news/2020/04/22/germany-got-right-fight-against-coronavirus/

Technical report

conclusion.

Over the forthcoming months it is likely that any 'flood' will be more of a 'trickle'. In this limited window of opportunity, FTTI protocols become viable, in the sense that detecting asymptomatic and infected individuals with a reasonable (e.g., 25%) efficacy would be sufficient to delay the re-emergence of the virus—a re-emergence that rests on, and only on, a slow loss of immunity. In short, FTTI will work after the first wave, even if it was not logistically viable before the first wave: if we lose collective immunity over a period of many months or years, then a second wave can be deferred or eluded, via FTTI, saving thousands of lives. This is under the proviso that a second wave can be pushed sufficiently far into the future, where it is rendered innocuous by vaccination or other interventions.

Clearly, this narrative depends on the acquisition of (herd) immunity following the first wave and its subsequent loss due to population fluxes, geo-social, and serological factors. The big question at the moment is whether the first wave has induced a sufficient level of herd immunity to open the window of opportunity for tracking and tracing. Although there are no current data for the UK, early studies in Germany speak directly to this issue. Figure 12 reproduces the hidden states in Figure 10, with a focus on infection status. The prevalence of immunity is shown as a yellow line. At the peak of the first wave (shortly after 60 days) the inferred level of immunity is 15%. This is the level of immunity estimated empirically in provisional reports of a serological study of people living in a region near Bonn (Streeck et al., 2020)[17]. Results of this sort are encouraging and endorse the inferences afforded by dynamic causal modelling. Perhaps more importantly, over the next few weeks more serological studies will become available[18] and we will know with greater certainty whether the above narrative is licensed by empirical data. If not, these data can be assimilated into the model to update our (Bayesian) beliefs about what has happened, what will happen and what could happen.

---

[17] Preliminary results from a town of about 12,500 in Heinsberg—a region in Germany that had been hit hard by COVID-19—suggest a seropositive prevalence of 14% of 1007 people assessed between March 30 and April 6, 2020.

[18] https://docs.google.com/spreadsheets/d/17Tf1Ln9VuE5ovpnhLRBJH-33L5KRaiB3NhvaiF3hWC0/edit#gid=0.

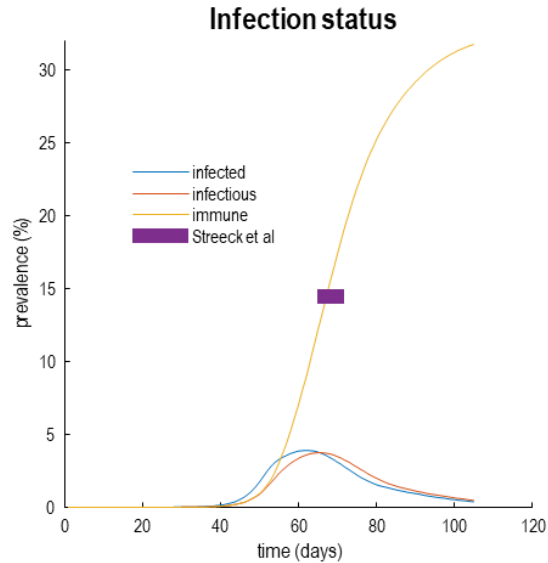

**FIGURE 12**

**Herd immunity**: This figure reproduces the results of Figure 10 but with a focus on infectious status. Here, the estimated prevalence of infected people, contagious (*infectious*) people and immune people are shown as a function of time since the onset of the outbreak (25th of January 2020). The purple bar indicates the level of estimated seroprevalence in a region in Germany as reported in (Streeck et al., 2020). This matches the predictions of the model; namely, a 15% herd immunity at the peak of the infection. According to this analysis, this level of herd immunity is contextualised by the proportion (about 42%) of the population that are susceptible to infection and subsequent morbidity.

## Limitations

There are clearly many limitations of the dynamic causal modelling above. Perhaps the most obvious is using a model of a single region to account for the spread of the virus over the four Nations of the United Kingdom—and the Upper Tier Local Authorities comprise each region. Multi-region versions of the current model have been implemented to understand the geographical spread of the outbreak in terms of regional connectivity and population fluxes (Friston et al., 2020b). This kind of fine-grained modelling speaks to a wave of infection emanating from an epicentre (e.g., London) spreading—much like a wildfire—across the country to reach distant regions, from Aberystwyth to Gateshead, from the Isle of Wight to Shetland. In short, pooling data from all four Nations affords a more precise estimate of temporal dynamics (i.e., the focus of this report) at the expense of neglecting regional waves and spatiotemporal dynamics. This may be especially relevant when considering the constitution of *resistant* individuals in the population: this cohort was introduced to explain data from the United States (Friston et al., 2020b). In other words, a proportion of resistant individuals was necessary to explain early mortality rates (and new cases), given the known population size and inferred characteristics of viral transmission. Intuitively, one could regard this resistant proportion as 'dark matter' in the universe that cannot be seen but is necessary to explain astronomical data[19].

So, what is the composition of this 'dark matter'? One could think of resistant individuals in terms of host specific factors: namely, a reduced propensity to contract the virus and participate in its transmission. This would be like assuming a certain proportion of people—e.g., in a household or care home—cannot play host to the virus, even when exposed. Alternatively, resistance could be geographical in nature; in the sense that some people are geographically insulated from exposure. Crucially, this constituent of resistance will be time-dependent if the virus is spreading towards a community that has yet to be exposed. So, what are the implications for such communities? On the current analysis, it suggests that if FTTI was rolled out across the four Nations, it could defer the first wave—in a previously unaffected region—indefinitely. For example, while London was deferring its second wave, Aberystwyth could defer its first.

This raises important questions about regions that have and have not experienced a first wave. At first glance, it may seem natural to reduce connectivity and population flux into hitherto unaffected regions. However, there are other hypotheses. For example, in multiregional modelling of the United States (Friston et al., 2020b), increasing the commuter traffic between two States sometimes improved predicted mortality. One intuition behind these paradoxical effects could be a protective effect of importing people with immunity from other regions; much like suppressing chain reactions with graphite in early nuclear reactors (e.g., doubling the number of medics in Aberystwyth with secondments from London could reduce the exposure of patients to contagion). These are speculations that could, in principle, be evaluated with sufficiently refined dynamic causal models.

# Appendices

## Modelling self-isolation

Equipped with an extra location (*isolation*) state enables one to distinguish between simply staying at home or being out and about (i.e., at *work*). These states are rough approximations to the different kinds of environment we find ourselves in and are used to differentiate the number of contacts that could potentially transmit the virus from one person to another. When considering the parameterisation of population dynamics, in terms of being in a particular state, one has to parameterise the time spent in that state, in relation to the probability of leaving or entering it. In this instance, the probability of entering self-isolation is unity when you develop symptoms or submit to PCR testing. You then remain in that state for seven days, unless you receive notification that you have tested negative. While in this state, you can neither infect nor be infected by anybody else. Mathematically, this can be parameterised as follows, where $\tau_{iso}$ is seven days (conditional on not being in critical care or the morgue)

---

[19] Note that the posterior estimates of the proportion of people who are resistant was higher than the prior (one third) for both the UK and Germany.

$$P(isolation_{t+1} | isolation_t) = \begin{cases} 1 & \text{if symptomatic, positive or waiting} \\ 0 & \text{if negative and asymptomatic} \\ \exp(-1/\tau_{iso}) & \text{otherwise} \end{cases} \quad (1.1)$$

Clearly, the transition from one *location* state to another now depends upon the *testing* factor. If you are *waiting* for a test, you move into *isolation* and if you are *negative* you leave. It is this conditional dependency between the factors that mediates the efficacy of FTTI.

## Modelling FTTI

The parameterisation of FTTI and other *testing* state transitions is a bit more delicate. This is because there are several reasons you might be tested that depend on several factors. In this model, we parameterise testing by first establishing a time-dependent probability of being tested on any given day. If we now condition the probability of being tested on whether or not you are infected, we need to parameterise the potential bias towards testing people who are infected. This means that the total probability of being tested is a weighted average of two probabilities, parameterised as follows (Please see Table 1 for a description of the parameters):

$$P(tested) = P(tested | infected)P(infected) + P(tested | not\ infected)P(not\ infected)$$
$$\Rightarrow$$
$$P_{tes} = P(tested | infected) = \theta_{tes} P_{sen} \quad (1.2)$$
$$P_{sen} = P(tested | not\ infected) = \frac{P(tested)}{\theta_{tes} P(infected) + 1 - P(infected)}$$
$$\theta_{tes} = \frac{P(tested | infected)}{P(tested | not\ infected)}$$

where the probability of being tested has three components:

$$P(tested) = \theta_{bas} + \theta_{sen} P(infected) + \theta_{exp} P(immune) \quad (1.3)$$

The first is a *baseline* probability. For the UK analyses, this parameter was proportional to the total number of tests up until the present time, and the maximum number of tests in the future[20]. The second parameterises a *sensitivity* to the prevalence of infection in the community and increases with infection rates. The final term is a *sustained* response following onset of the first wave. This sustained component is necessary to model countries that ramp up their testing capacity during the first peak. Here, we use the level of immunity as a proxy for a gently declining function of the cumulative number of infected people. Each of these terms has a parameter enabling one to fit a time-dependent probability of being tested. Finally, we have to consider targeted testing of individuals who have been identified as having been in contact with an infected individual. This affords an enhanced probability of testing if, and only if, you are infected and asymptomatic. By adding this probability to the probability of being tested when infected, we supplement general screening with a FTTI parameter as follows:

$$P(tested \mid infected, asymptomatic) = P_{tes} + \theta_{ttt}(1 - P_{tes}) \quad (1.4)$$

---

[20] This means that the total tests for the UK entered the analysis twice. First, when normalised to their maximum, they provide empirical prior constraints on the ramp up of baseline testing in the UK. They then enter unnormalised as data to be explained.

The ensuing parameter is simply the efficacy or extra probability that I will be tested if I am infected and asymptomatic. If it were possible to trace and test everybody who has been exposed and contracted the virus prior to developing symptoms, this efficacy will be one. In the absence of any targeted testing efficacy will be zero. A priori, the efficacy was set to very low levels of one in 10,000 people, per day.

## Effective reproduction rate

The effective reproduction rate is a fundamental epidemiological constant that provides a useful statistic that reflects the exponential growth of the prevalence of infection. There are several ways in which it can be formulated and estimated. For our purposes, we can generate an instantaneous reproduction rate directly from the time varying prevalence of infection as follows:

$$\begin{aligned} R_t &= \exp(K_t \cdot \tau_{con}) \\ K_t &= \ln \frac{P(infected_{t+1})}{P(infected_t)} = \frac{\ln(2)}{T_d} \end{aligned} \quad (1.5)$$

These expressions show that the reproduction rate reflects the growth of the (logarithm of) proportion of people infected—and the period of being infectious. This number is formally related to the doubling time $T_d$. Note that the reproduction rate is not an estimate in this scheme: it is an outcome that is generated by the latent causes or hidden states inferred by inverting (i.e., fitting) the model to empirical timeseries.

## Software note

The figures in this report can be reproduced using annotated (MATLAB/Octave) code available as part of the free and open source academic software SPM (https://www.fil.ion.ucl.ac.uk/spm/), released under the terms of the GNU General Public License version 2 or later. The routines are called by a demonstration script that can be invoked by typing >> DEM_COVID_T at the MATLAB prompt. At the time of writing, these routines are undergoing software validation in our internal source version control system—that will be released in the next public release of SPM (and via GitHub at https://github.com/spm/). In the interim, please see https://www.fil.ion.ucl.ac.uk/spm/covid-19/.

The data used in this technical report are available for academic research purposes from the COVID-19 Data Repository by the Center for Systems Science and Engineering (CSSE) at Johns Hopkins University, hosted on GitHub at https://github.com/CSSEGISandData/COVID-19, and from Tom White on GitHub at https://github.com/tomwhite/covid-19-uk-data.

Technical report

## Acknowledgements

This work was undertaken by members of the Wellcome Centre for Human Neuroimaging, UCL Queen Square Institute of Neurology. The Wellcome Centre for Human Neuroimaging is supported by core funding from Wellcome [203147/Z/16/Z]. A.R. is funded by the Australian Research Council (Refs: DE170100128 and DP200100757). A.J.B. is supported by a Wellcome Trust grant WT091681MA. CL is supported by an MRC Clinician Scientist award (MR/R006504/1).

The authors declare no conflicts of interest.## References